\documentclass[preprint,12pt]{elsarticle}

\usepackage{amssymb}
\usepackage{amsmath}

\journal{Acta Astronautica}

\begin{document}

\begin{frontmatter}



\title{Orbit-localised thermosphere density prediction using a Kalman filter based calibration of empirical models}


\author[inst1]{George Bowden}

\affiliation[inst1]{organization={School of Engineering and Information Technology},
            addressline={The University of New South Wales}, 
            city={Canberra},
            postcode={2612}, 
            state={ACT},
            country={Australia}}

\begin{abstract}
Accurate estimation of thermosphere mass density is critical to determining how satellite orbits evolve over time and thus to planning and managing space missions. Empirical thermosphere models are commonly employed for this purpose, but have substantial uncertainties. In this work, a Kalman filter method for calibrating these models along a particular satellite's trajectory is described. This method was applied to calibrate the NRLMSISE-00, JB2008, and DTM-2020 models with respect to densities measured by the Swarm-C satellite. Substantial improvements in root mean squared density residuals were obtained using the technique when compared with either uncalibrated model output or calibration using a linear regression on previous data. Further improvement was obtained by combining estimates from different models using a best linear unbiased estimator method.

\end{abstract}


\begin{highlights}
\item A method is described for estimating the time-varying linear calibration of thermosphere density models for a particular satellite using a Kalman filtering algorithm.
\item The method is applied to orbit means and principal components of densities from precise orbit determination of the Swarm C satellite.
\end{highlights}


\end{frontmatter}


\section{Introduction}
\label{sec:introduction}

Aerodynamic drag is the greatest source of uncertainty for the determination and propagation of satellite trajectories in Low Earth Orbit (LEO). This force arises from interaction with the thermosphere, the atmospheric layer from $\sim 90$ to $600$ km altitude. Accurately estimating satellite drag is a key challenge for mission planning, lifetime estimation, reentry prediction, collision avoidance, and propulsionless manoeuvring. Satellite drag, $F_D$, can be expressed in terms of atmospheric density, $\rho$, drag coefficient, $C_D$, speed relative to the atmosphere, $v$, and reference area, $A$,
\begin{equation}
F_D = \frac{C_D \rho v^2 A}{2} .
\end{equation}
For a particular satellite, $F_D$ may be inferred from on-board accelerometer measurements or orbit determination from some combination of radar, laser ranging, and Global Navigation Satellite System (GNSS) measurements.

Significant uncertainty typically exists regarding thermospheric density and satellite aerodynamic properties \cite{Vallado2014}. Drag coefficient is determined by complex flow physics and therefore challenging to compute accurately. It is dependent on often complicated satellite geometries, gas-surface interactions, and thermosphere composition, though is not sensitive to density, velocity, and length scale. Thermosphere densities vary due to fluctuations in incoming extreme ultraviolet (EUV) radiation, electrical energy and energetic particles from the solar wind, and waves and instabilities propagating from the lower atmosphere \cite{Emmert2015}. Empirical thermosphere models incorporate the former two influences through parameterisations based on climatological data. However, their accuracy is limited by the relative sparseness of underlying data and unmodelled physics.

An alternative to applying global thermosphere models to individual satellites is to obtain orbit-localised density estimates for those satellites based on their previous density measurements. Here density estimates are restricted to individual satellite trajectories and based on prior density measurements from the satellites. An example of such an approach, employing a linear autoregressive filter to forecast density data from the CHAllenging Minisatellite Payload (CHAMP) satellite, was described by Stastny et al. \cite{Stastny2009}. Similarly, P\'{e}rez et al. detailed the use of feed-forward time-delay artificial neural networks (ANNs) to predict density for CHAMP \cite{Perez2014}. Solar and geomagnetic index data were included as inputs to incorporate space weather effects. These computationally inexpensive models could feasibly be implemented as part of on-board orbit propagation.

In addition to the aforementioned work, P\'{e}rez and Bevilacqua investigated the use of ANNs for calibrating density predictions from empirical thermosphere models \cite{Perez2015}. Their networks were trained and validated on the Gravity Recovery and Climate Experiment (GRACE) A data, though tested on CHAMP data. The influence of uncertainty in the drag coefficient of a satellite on its orbit determination and propagation could be reduced by calibrating density estimates for that particular satellite. This calibration would help ensure that density and drag coefficient estimates together estimate drag accelerations accurately, despite there being significant uncertainty in the former quantities individually.

Here a Kalman filter based method for calibrating thermosphere density estimates from empirical models along the trajectory of a particular satellite is described and evaluated. Such a method can be used with irregularly spaced and delayed density data, by contrast with prior orbit-localised methods. The method can account for data with non-uniform uncertainties. It also has the advantage of providing uncertainties associated with its estimates. Furthermore, the filter parameters indicate how rapidly the relationship between empirical model estimates and satellite drag changes.

In this Kalman filter implementation, the calibration is assumed to be linear with temporal evolution described by a stochastic process. It is iteratively updated based on measured orbit densities. This technique is applied to historical data for the Swarm-C satellite, one component of a constellation studying Earth's magnetic field \cite{Friis2008}.

\section{Empirical thermosphere models}
\label{sec:empirical_thermosphere_models}

Empirical thermosphere models represent the dependence of density on parameters such as altitude, local time, latitude, day of year, solar EUV flux, and geomagnetic activity. They are considered climatological, being based on average observed behaviour, though usually incorporate certain physics-based constraints. The development of such models is discussed in review articles by Vallado and Finkleman \cite{Vallado2014} and Emmert \cite{Emmert2015}. Estimates vary between models due to use of different thermosphere data-sets in their derivation, different space weather indices as input, and different fitting functions to represent dependence on time and location. Three empirical models, summarised below, are considered in this study. The model implementations used here are available online \cite{Mahooti2019,Mahooti2018,SWAMI2021}.

\subsection{NRLMISE-00}
\label{sub:msis}
The Naval Research Laboratory Mass Spectrometer Incoherent Scatter radar Extended 2000 model (NRLMSISE-00) \cite{Picone2002} belongs to the MSIS family of empirical thermosphere models \cite{Hedin1977a,Hedin1977b}. This model provides mass density, number densities of major thermosphere species, and neutral temperature. It was derived from a combination of satellite mass spectrometer (MS), incoherent scatter radar (ISR), satellite accelerometer, orbit determination, occultation, and sounding rocket data. Solar EUV variation is included through daily and $81$-day averaged $F_{10.7}$ solar flux indices (ground-based radio proxies for EUV radiation) and geomagnetic activity is included through a history of $3$-hourly planetary $a_p$ indices.

\subsection{JB2008}
\label{sub:jb2008}
The Jacchia-Bowman 2008 model (JB2008) is based on satellite orbit determination and accelerometer data \cite{Bowman2008}. The model computes mass density and temperature. As well as the $F_{10.7}$ index, the model incorporates $S_{10.7}$, $M_{10.7}$, and $Y_{10.7}$ indices derived from satellite-based EUV, ultraviolet, and X-ray measurements respectively.  Geomagnetic activity is represented through estimated changes in exosphere temperature, which are derived from the Disturbance Storm Time ($Dst$) index.

\subsection{DTM-2020}
\label{sub:dtm}
The Drag Temperature Model 2020 (DTM2020) \cite{Bruinsma2021a} was developed from the DTM family of empirical thermosphere models \cite{Barlier1978,Bruinsma2015}. These models are fitted to satellite MS, ISR, satellite accelerometer, orbit determination, and optical spectrometer measurements. The version of the model in this study takes as input daily and $81$-day averaged $F_{10.7}$ to represent EUV variation and $3$-hour delayed and $24$-hour averaged $K_p$ indices to represent geomagnetic activity.

\section{Kalman filter approach}
\label{sec:kalman_filter_approach}
The Kalman filter is a widely-used algorithm for recursively estimating the state of a system based on noisy measurements \cite{Kalman1960}. This filter determines the most likely state of a system with additive Gaussian measurement and process noise. It is applicable to smoothing, filtering, and prediction problems. Here, the Kalman filter is used to estimate coefficients for calibration of empirical model orbit mean thermosphere mass density encountered by a satellite. Scores for each thermosphere mass density principal component can be calibrated using a similar method. Combining these data allow estimation of thermosphere mass density at specific locations on the satellite orbit.

\subsection{Transition model and measurements}
\label{sub:transition_and_measurements}

Assume that the mean thermosphere mass density over an orbit has an expectation value which is linearly related to that predicted by an empirical model, $\rho_m$ (i.e. $\left < \rho \right > = m \rho_m + c $ where $m$ and $c$ are constants). Furthermore, assume that the measured value is perturbed by a stochastic component with a Gaussian distribution $p \left ( \epsilon_\rho \right ) \sim  \mathcal{N} \left ( 0 , R \right )$. This term represents both measurement noise and the effect of transient perturbations excluded from climatological models, such as atmospheric waves. Thus, the measured mean density can be expressed as
\begin{equation}
    \label{eq:lin_rho}
    \rho = m \rho_m + c + \epsilon_\rho .
\end{equation}

A state vector can be constructed from the coefficients in Eq.~\ref{eq:lin_rho} for time-step $k$,
\begin{equation}
    \mathbf{x}_k = \left[ \begin{array}{c}
m_k \\
c_k \end{array} \right] .
\end{equation}
Temporal evolution is modelled as a Gaussian random walk process,
\begin{equation}
    \mathbf{x}_{k+1} = \mathbf{F}_k \mathbf{x}_k + \mathbf{w}_k
\end{equation}
where state transition matrix $\mathbf{F}_k$ is the identity matrix and $p \left ( \mathbf{w}_k \right ) \sim \mathcal{N} \left ( 0 , \mathbf{Q}_k \right ) $ where $\mathbf{Q}_k = \left ( t_{k+1} - t_k \right ) \mathbf{M} $. Eq.~\ref{eq:lin_rho} can be rewritten in matrix form to represent an observation at time $k$,
\begin{equation}
    \mathbf{z}_k = \mathbf{H}_k \mathbf{x}_k + \mathbf{v}_k ,
\end{equation}
where $\mathbf{z}_k = \rho$, $\mathbf{H}_k = \left [ \begin{array}{c c}
\rho_m & 1 \end{array} \right ]$, and $p \left ( \mathbf{v}_k \right ) \sim  \mathcal{N} \left ( 0 , R \right )$.

\subsection{Filter algorithm}
\label{sub:filter_algorithm}
A Kalman filter provides the optimal (maximum likelihood) estimate of the updated state, $\hat{\mathbf{x}}_k$, and its associated covariance matrix, $\mathbf{P}_k$, based on prior values, $\hat{\mathbf{x}}_k^-$ and $\mathbf{P}_k^-$, and observation $\mathbf{z}_k$. First, the state and covariance are updated based on the density measurement at time $k$,
\begin{equation}
    \hat{\mathbf{x}}_{k} = \hat{\mathbf{x}}_{k}^- + \mathbf{K}_k \left ( \mathbf{z}_k - \mathbf{H}_k \hat{\mathbf{x}}_k^- \right )
\end{equation}
\begin{equation}
    \mathbf{P}_k = \left ( \mathbf{I} - \mathbf{K}_k \mathbf{H}_k  \right ) \mathbf{P}_k^- ,
\end{equation}
where $\mathbf{K}_k$ is the optimal Kalman gain,
\begin{equation}
    \mathbf{K}_k = \mathbf{P}_k^- \mathbf{H}_k^T \left ( \mathbf{H}_k \mathbf{P}_k^- \mathbf{H}_k^T + \mathbf{R}_k \right )^{-1} .
\end{equation}
The state is then advanced in time
\begin{equation}
    \hat{\mathbf{x}}_{k+1}^- = \mathbf{F}_k \hat{\mathbf{x}}_k
\end{equation}
\begin{equation}
    \mathbf{P}_{k+1}^- = \mathbf{F}_k \mathbf{P}_k \mathbf{F}_k^T + \mathbf{Q}_k .
\end{equation}
This provides prior values for the next measurement update, so the process can be repeated.

In practice, for nowcasting and forecasting such a calibration must be used some time period after the data from which it was derived. If this is done $n$ time steps ahead, the estimated density and measurement uncertainty are found to be 
\begin{equation}
    \tilde{\rho}_{k+n} = \mathbf{H}_{k+n} \left ( \prod_{i=1}^n \mathbf{F}_{k+i-1} \right ) \hat{\mathbf{x}}_k
\end{equation}
and
\begin{multline}
    \tilde{\sigma}_{k+n} = \mathbf{H}_{k+n}  \left ( \left ( \prod_{i=1}^n \mathbf{F}_{k+i-1} \right ) \mathbf{P}_k \left ( \prod_{i=1}^n \mathbf{F}_{k+i-1} \right )^T \right . \\
    + \left . \sum_{i=1}^n \left ( \prod_{j=i+1}^n \mathbf{F}_{k+j-1} \right ) \mathbf{Q}_{k+i-1} \left ( \prod_{j=i+1}^n \mathbf{F}_{k+j-1} \right )^T \right ) \mathbf{H}_{k+n}^T + \mathbf{R}_{k+n} .
\end{multline}
For the system described in Subsection~\ref{sub:transition_and_measurements} these equations reduce to
\begin{equation}
    \tilde{\rho}_{k+n} = \mathbf{H}_{k+n} \hat{\mathbf{x}}_k
\end{equation}
and
\begin{equation}
    \label{eq:var_est}
    \tilde{\sigma}_{k+n} = \mathbf{H}_{k+n}  \left ( \mathbf{P}_k + n\mathbf{M} \right ) \mathbf{H}_{k+n}^T + R .
\end{equation}

\subsection{Parameter estimation}
\label{sub:parameter_estimation}
The filter and process noise parameters, $R$ and $\mathbf{M}$, are challenging to estimate on theoretical grounds. Both short time-scale thermosphere density perturbations and random errors in density measurements are expected to contribute significantly to $R$. These may not be independent, given that both can be associated with changes in thermosphere composition. Moreover, the somewhat abstract nature of a state vector formed from calibration coefficients makes it difficult to relate $\mathbf{M}$ to known quantities. Consequently, the best approach is to determine $R$ and $\mathbf{M}$ based on filter performance for a set of data. These are referred to as `training' data, by analogy with machine learning.

A maximum likelihood method is used to compute $R$ and $\mathbf{M}$. The marginal probability density $p \left ( R , \mathbf{M} | \rho_i \right )$ is maximised, where $\rho_i$ is the set of $m$ density measurements in the training data. Following the derivation of Eq.~14 of Mehra's paper on adaptive filtering \cite{Mehra1972}, the likelihood estimate can be obtained by maximising
\begin{multline}
    L \left ( R , \mathbf{M} \right ) = -\frac{1}{2} \sum_{i=1}^m \left \{ \frac{\left ( \rho_i - \mathbf{H}_i \hat{\mathbf{x}}_{i-n} \right )^2}{\mathbf{H}_i  \left ( \mathbf{P}_{i-n} + n\mathbf{M} \right ) \mathbf{H}_i^T + R} \right . \\
    \left . + \ln \left ( \mathbf{H}_i  \left ( \mathbf{P}_{i-n} + n\mathbf{M} \right ) \mathbf{H}_i^T + R \right ) \right \} .
\end{multline}
Here it is assumed that there is no prior information regarding the values of $R$ and $\mathbf{M}$, which would otherwise result in the addition of the term $\ln{\left ( p \left ( R , \mathbf{M} \right ) \right )}$ to the right hand side of the above equation. 

Unconstrained numerical optimisation is performed using the log-Cholesky parametrisation $M = L L^T$ where $L$ is a real lower triangular matrix \cite{Pinheiro1996}. The optimisation parameters are the logarithms of on-diagonal elements of $L$ and values of off-diagonal elements. Doing so ensures that $M$ is a positive definite symmetric matrix, which are essential properties of a realistic covariance matrix. Similarly, using the optimisation parameter $\ln{R} $ to represent $R$ ensures that this parameter has a realistic positive value.

\subsection{Combining density estimates}
\label{sub:combining_density_estimates}
Using the above, Kalman filters can be obtained for calibrating various empirical thermosphere models. This leads to the question of how to best combine their density estimates. Residuals in these predictions cannot be considered independent, given the common contributions of measurement errors and non-climatological perturbations. The best linear unbiased estimate based on correlated estimates is \cite{lyons1988}
\begin{equation}
\tilde{\rho}_c = \boldsymbol{\alpha}^T \tilde{\boldsymbol{\rho}} ,
\end{equation}
where the weight vector is 
\begin{equation}
    \boldsymbol{\alpha} =  \frac{\mathbf{K}_{\boldsymbol{\epsilon}_\rho \boldsymbol{\epsilon}_\rho}^{-1} \mathbf{u}}{\mathbf{u}^T \mathbf{K}_{\boldsymbol{\epsilon}_\rho \boldsymbol{\epsilon}_\rho}^{-1} \mathbf{u}}
\end{equation}
with $\mathbf{u}$ being a vector in which each element is unity. The variance is estimated by
\begin{equation}
    \sigma_{\rho_c}^2 = \boldsymbol{\alpha}^T \mathbf{K}_{\boldsymbol{\epsilon}_\rho \boldsymbol{\epsilon}_\rho} \boldsymbol{\alpha} .
\end{equation}
In these equations, $\tilde{\boldsymbol{\rho}}$ is a vector of calibrated orbit mean densities and $\mathbf{K}_{\boldsymbol{\epsilon}_\rho \boldsymbol{\epsilon}_\rho}$ is the covariance matrix for residuals $\epsilon_\rho$.

While Eq.~\ref{eq:var_est} provides an estimate of the variance for the estimates derived from each empirical model, it does not provide a straightforward method of computing off-diagonal terms of $\mathbf{K}_{\boldsymbol{\epsilon}_\rho \boldsymbol{\epsilon}_\rho}$. Therefore, $\mathbf{K}_{\boldsymbol{\epsilon}_\rho \boldsymbol{\epsilon}_\rho}$ is instead estimated based on residuals for the training data. As $\mathbf{P}_k$ is found to stabilise very quickly for constant $\mathbf{Q}_k$ and $\mathbf{R}_k$ in practice, the assumption of constant $\mathbf{K}_{\boldsymbol{\epsilon}_\rho \boldsymbol{\epsilon}_\rho}$ is reasonable.

\section{Example case: Swarm-C densities}
\label{sec:example_case}

\subsection{Swarm-C precise orbit determination densities}
\label{sub:pod_densities}
The thermosphere mass density data used in this study were values published by ESA derived from Swarm-C GNSS measurements using Precise Orbit Determination (POD) \cite{VanIJssel2020}. This entailed estimation of aerodynamic accelerations using an extended Kalman filter method with a force model incorporating gravitational and radiation forces. Densities were calculated from these accelerations using direct simulation Monte Carlo simulation to compute $C_D$. Thermosphere composition was taken from NRLMSISE-00 and a constant accommodation coefficient (ratio of energy exchanged in gas-surface interaction to its thermodynamic maximum) assumed. The fastest density variations this method captured were a factor of $4$-$5$~times shorter than the orbital period, corresponding to a resolution of about $80^{\circ}$ in argument of latitude, $u$. However, the published data had a $30$~s sampling rate.

\subsection{Data selection}
\label{sub:selection}
\begin{figure}
    \centering
        \includegraphics{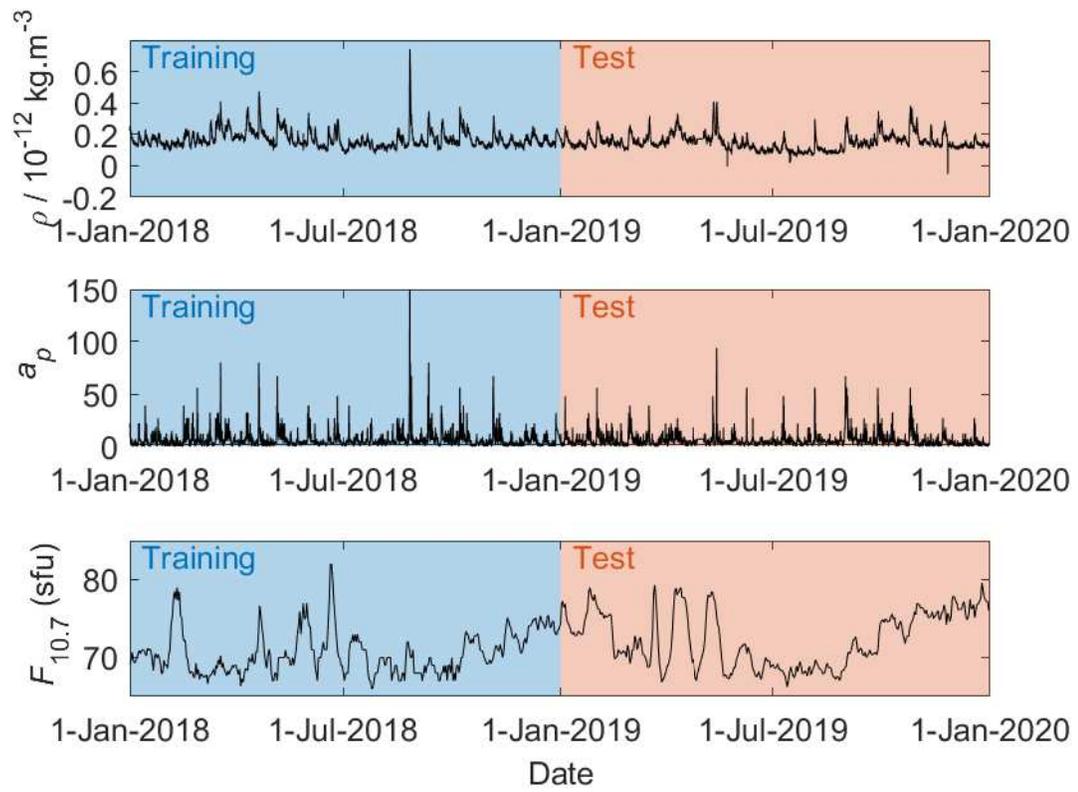}
    \caption{Swarm-C orbit mean density, $\rho$, geomagnetic activity, $a_p$, and solar flux index, $F_{10.7}$, for $2018$ and $2019$.}\label{fig:input_summary}
\end{figure}

Density measurements from 2018 and 2019 constituted the training and test data sets respectively. These data coincided with the solar minimum at the end of Solar Cycle 24. During this time period, Swarm-C had an altitude of approximately $450$~km and orbital inclination of $87.4^{\circ}$. Slow nodal precession occurred such that local time of the ascending node (LTAN) covered the full $24$~hour range over a period of approximately $9$~months. Training and test orbit mean density data are plotted along with space weather indices in Fig.~\ref{fig:input_summary}.

\subsection{Data processing}
\label{sub:processing}
For each orbit during the training and test periods, Swarm-C POD density data were interpolated to a grid with spacing $1^{\circ}$ in satellite argument of latitude, $u$. Orbits with missing data or erroneous values were excluded from the analysis. Mean orbit densities were then computed averaging over $u$.

\begin{figure}
    \centering
        \includegraphics{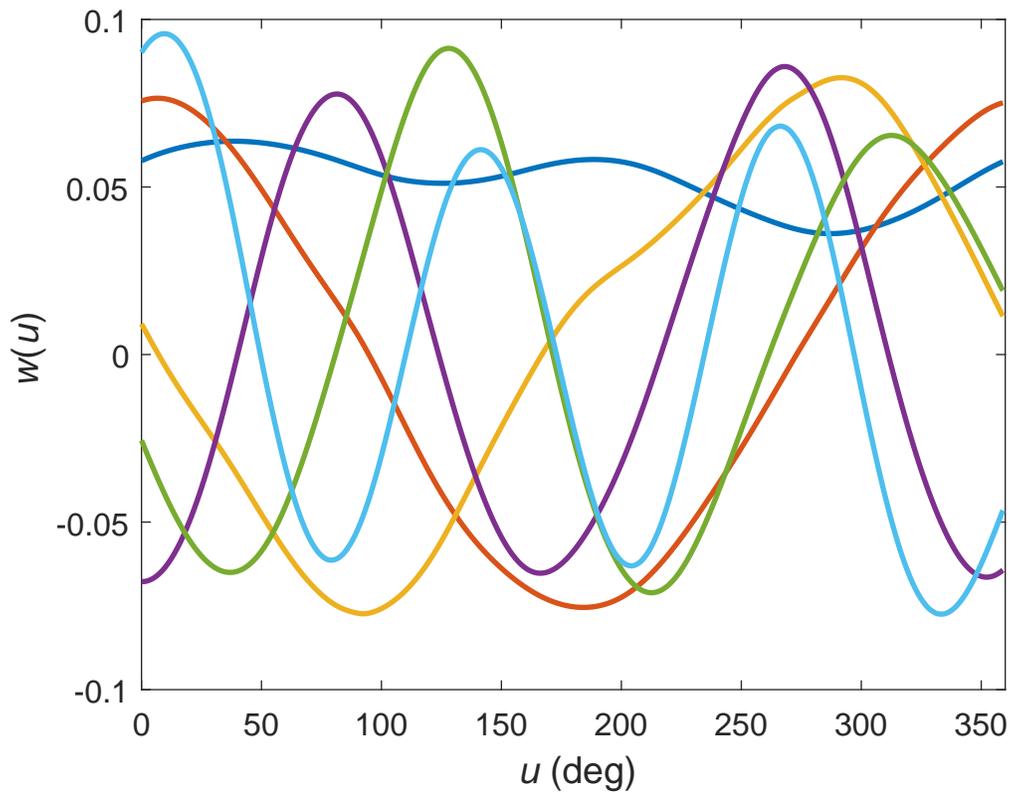}
    \caption{Principal components of Swarm-C precise orbit determination density as functions of argument of latitude, $u$, for $2018$. Coefficients for the first (blue), second (orange), third (yellow), fourth (purple), fifth (green), and sixth (cyan) components are plotted. Color available in online version.}\label{fig:pc_pod_density}
\end{figure}

\begin{figure}
    \centering
        \includegraphics{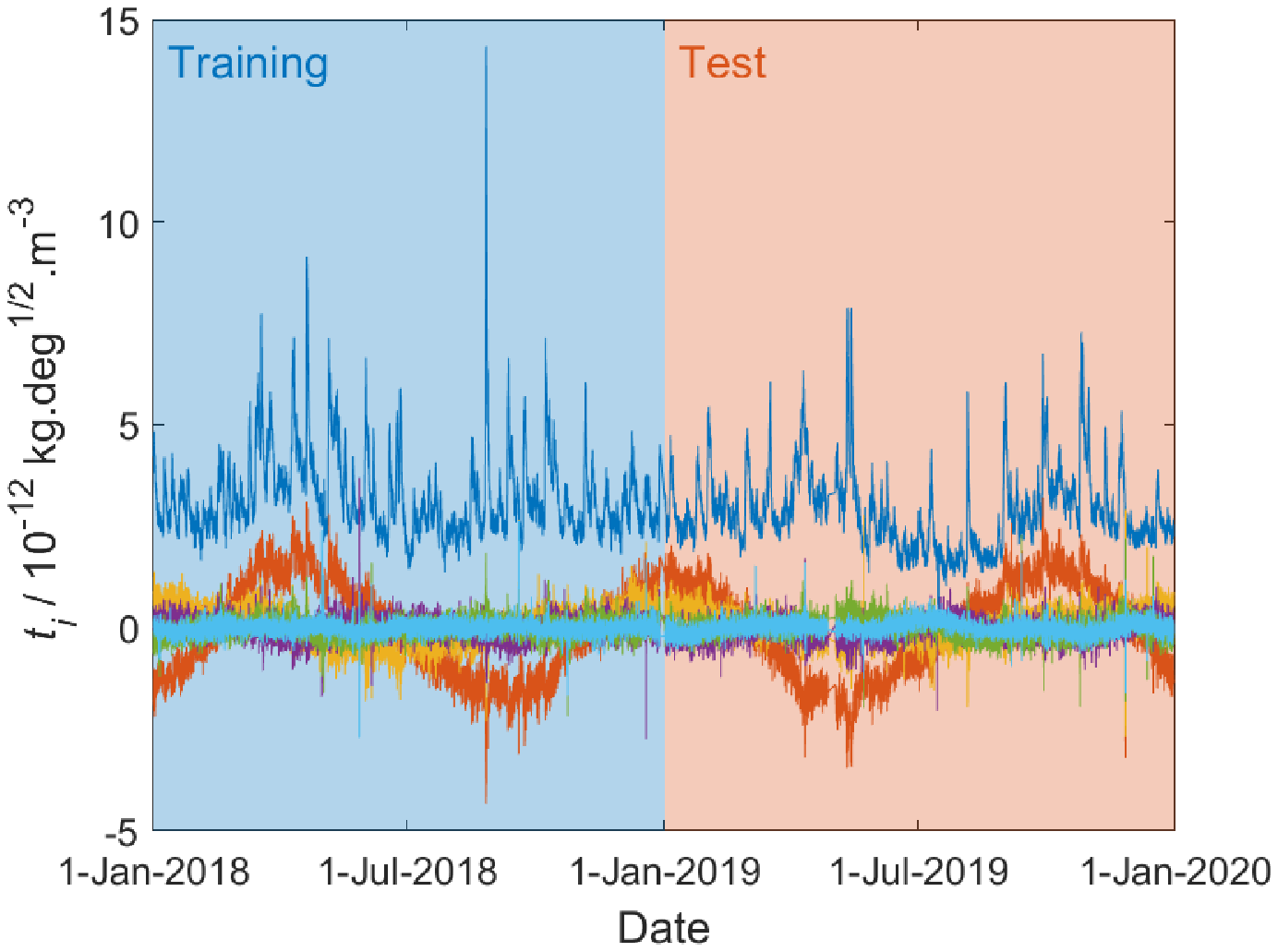}
        \caption{Scores for Swarm-C precise orbit determination density in $2018$ and $2019$, for the principal components plotted in Fig.~\ref{fig:pc_pod_density}. Scores for first (blue), second (orange), third (yellow), fourth (purple), fifth (green), and sixth (cyan) components are plotted.}\label{fig:pc_pod_scores}
\end{figure}

Principal component coefficients were computed for the training period data. This was done without the usual practice of subtracting mean values for each $u$, as means for a training period will not be applicable for different periods of the solar cycle. Consequently, the first principal component reflects these mean values in addition to changes driven by geomagnetic and solar activity. The first $6$ principal components, those accounting for more than $1\%$ variation in density, are plotted in Fig.~\ref{fig:pc_pod_density}. These were used to compute scores, $t_i$, for both the training and test data sets, plotted in Fig.~\ref{fig:pc_pod_scores}.

The first principal component largely represents the global mean density, with changes driven by variations in geomagnetic and solar activity. Peaks in its value correspond to times of elevated $a_p$ index in Fig.~\ref{fig:pc_pod_density}. By contrast, the second principal component correlates with density changes due to nodal precession. Fig.~\ref{fig:pc_pod_scores} indicates that the corresponding score varies with mean $0$ and period of about $9$~months. This function is continuous from $u = 360^{\circ}$ to $0^{\circ}$ and is approximately odd symmetric about $u = 90^{\circ}$, as expected for a circular polar orbit. The third component seems to reflect seasonal variation in the thermosphere itself. It has period $12$~months, is continuous from $u = 360^{\circ}$ to $0^{\circ}$ and is approximately odd symmetric about $u = 0^{\circ}$. The fifth and sixth principal components have large discontinuities from $u = 360^{\circ}$ to $0^{\circ}$ corresponding to changes in density during the time the satellite takes to complete an orbit.

\subsection{Orbit mean data}
\label{sub:orbit_mean}
The Kalman filter based calibration technique detailed in Section~\ref{sec:kalman_filter_approach} was applied to the orbit mean density data and estimates of the NRLMSISE-00, JB2008, and DTM-2020 models. For the purpose of this example, the NRLMSISE-00 and DTM-2020 empirical models were driven with the definitive solar flux and geomagnetic indices available from CLS (Collecte Localisation Satellites) \cite{CLS} and GFZ (German Research Centre for Geosciences) \cite{GFZ} respectively. The JB2008 model used data available from \cite{SET}.

The method was applied estimating the calibration coefficient $1$ and $3$~days ahead. This could represent the case of very short-term density prediction (nowcasting) where accurate and timely solar flux and geomagnetic indices are available but satellite-based density measurments are delayed. Alternatively, it can be considered a best case scenario for density forecasts where solar flux and geomagnetic indices can be predicted very accurately.

For comparison purposes, fixed calibration coefficients were obtained from linear least squares regression. These were computed separately from both the training and the test data. The former represented calibration of test data based solely on training data. The latter represented the best possible calibration with fixed coefficients for the test data. However, linear regression is clearly not a practical method of obtaining this calibration as it incorporates future data.

The root mean square (RMS) residuals for orbit mean densities estimated using the four calibration methods are tabulated in Tab.~\ref{tab:oa_results}. These may be compared with the mean value of $0.1496 \times 10^{-12}$ kg.m$^{-3}$ for the test data. The calibrated RMS residuals in the table can be compared with the corresponding values without calibration; for NRLMSISE-00, JB2008, and DTM-2020 these are $0.1720 \times 10^{-12}$ kg.m$^{-3}$, $0.0548 \times 10^{-12}$ kg.m$^{-3}$, and $0.0349 \times 10^{-12}$ kg.m$^{-3}$ respectively. Tab.~\ref{tab:oa_results} also includes residuals for results combined using the method detailed in Subsection~\ref{sub:combining_density_estimates}.

\begin{table}
\centering
\caption{RMS residuals for different orbit mean density calibration methods. All quantities are shown in $10^{-12}$ kg.m$^{-3}$ (without brackets) and ratio with the mean value (within brackets).}\label{tab:oa_results}
\begin{tabular}{ l c c c c }
\hline
 Calibration method  & NRLMSISE-00 & JB2008 & DTM-2020 & Combined \\
\hline
Linear regression \\ on training data & 0.0469 (0.31) & 0.0379 (0.25) & 0.0478 (0.32) & 0.0411 (0.27) \\
Linear regression \\ on test data & 0.0306 (0.20) & 0.0286 (0.19) & 0.0235 (0.16) & 0.0214 (0.14) \\
$1$-day offset \\ Kalman filter & 0.0279 (0.19) & 0.0278 (0.19) & 0.0247 (0.16) & 0.0211 (0.14) \\
$3$-day offset \\ Kalman filter & 0.0286 (0.19) & 0.0300 (0.20) & 0.0244 (0.16) & 0.0233 (0.16) \\
\hline
\end{tabular}
\end{table}

\begin{figure}
    \centering
        \includegraphics{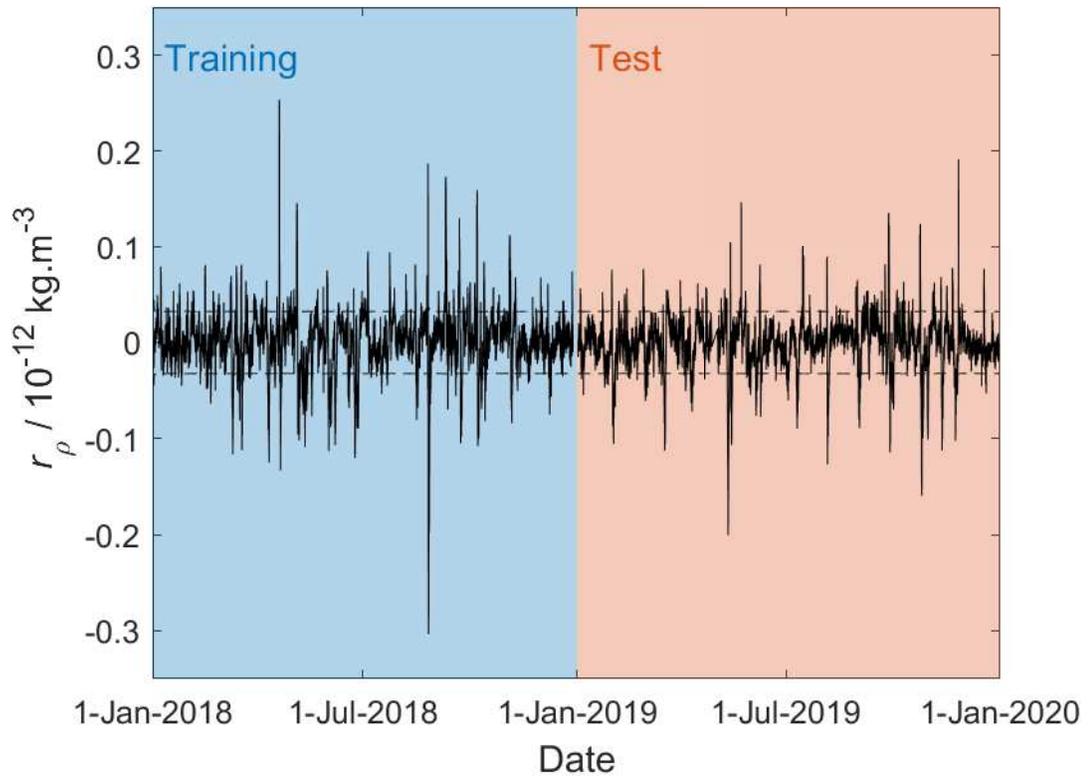}
    \caption{Mean orbit density residuals, $r_{\rho}$, for 1-day offset Kalman filter calibration of NRLMSIS-00 output (solid line). $\pm \sqrt{S'}$, the predicted standard error (dashed lines).}\label{fig:res_ic_msis}
\end{figure}

Residuals are plotted for estimation of the calibration coefficient for NRLMSIS-00 $1$~day ahead in Fig.~\ref{fig:res_ic_msis}. Also plotted for comparison is the predicted standard error for this quantity, $\sqrt{S'}$, which adopted a near constant value. The mean value of $\sqrt{S'}$ was $0.0337 \times 10^{-12}$ kg.m$^{-3}$, approximating the RMS value of the residual $0.0279 \times 10^{-12}$ kg.m$^{-3}$. Similar results were found for each model.

\begin{figure}
    \centering
        \includegraphics{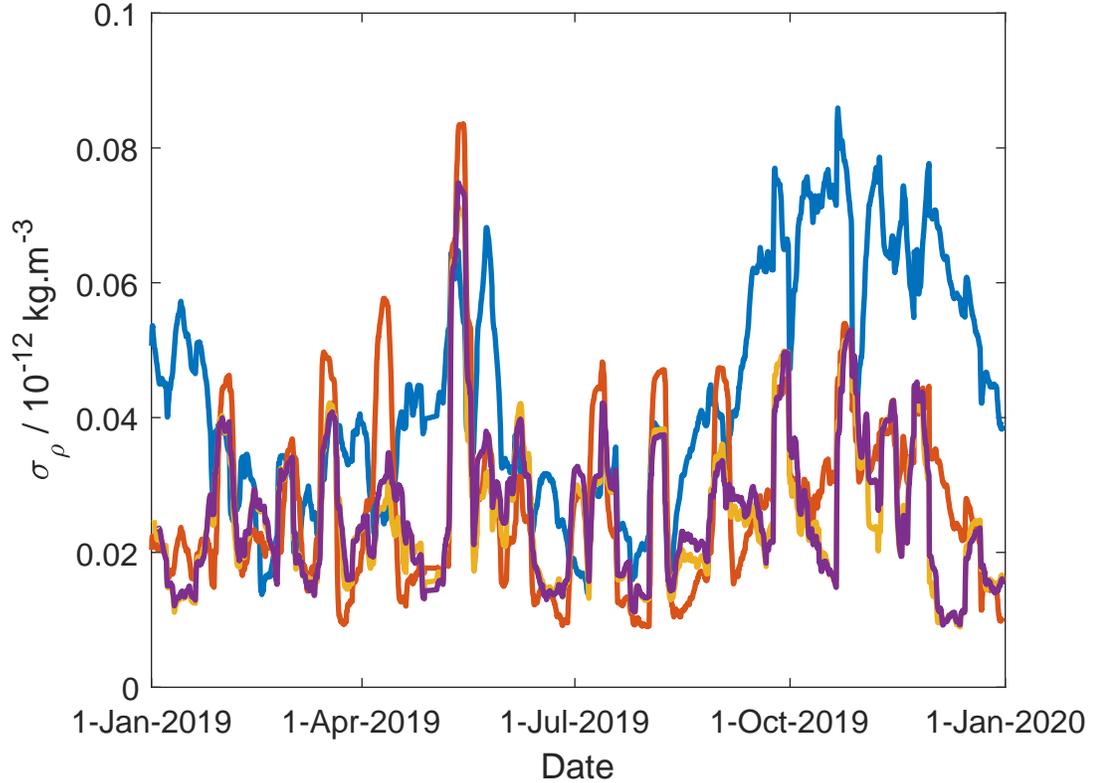}
    \caption{RMS mean orbit density residual values, taken over a moving window of 100 orbits ($\sim 1$~week). Plotted for NRLMSISE-00 training data linear regression (blue), test data linear regression (orange), 1-day offset Kalman filter calibration (yellow), and 3-day offset Kalman filter calibration estimates (purple).}\label{fig:movrmsres_oa_msis}
\end{figure}

\begin{figure}
    \centering
        \includegraphics{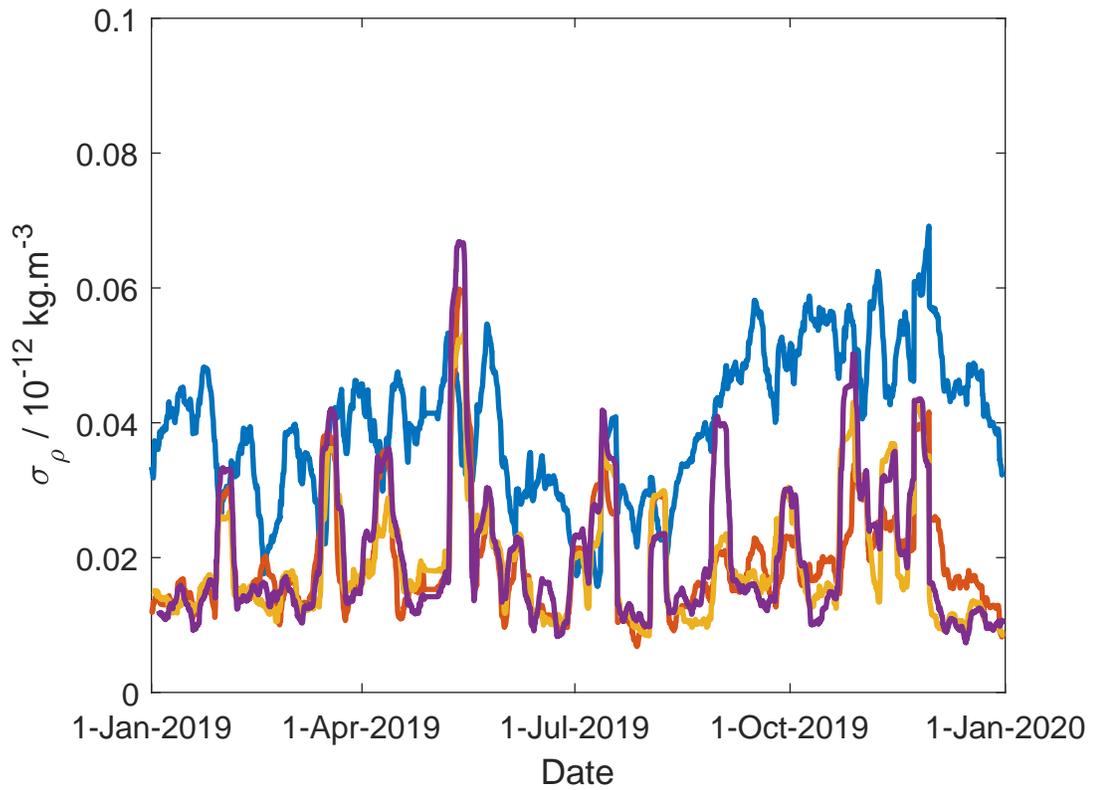}
    \caption{RMS mean orbit density residual values, taken over a moving window of 100 orbits ($\sim 1$~week). Plotted for combined training data linear regression (blue), test data linear regression (orange), 1-day offset Kalman filter calibration (yellow), and 3-day offset Kalman filter calibration estimates (purple).}\label{fig:movrmsres_oa_all}
\end{figure}

Temporal variation of the orbit mean density residuals is plotted in Fig.~\ref{fig:movrmsres_oa_msis} for NRLMSISE-00 and Fig.~\ref{fig:movrmsres_oa_all} for the combined output. These plots show RMS residuals computed over a moving window of $100$ orbits, omitting missing values. In each case, there are periods where the magnitude of residuals for the test data linear regression and Kalman filter calibrations sharply increased. Comparing the plots with Fig.~\ref{fig:input_summary}, these periods often coincided with increases in density and relatively high geomagnetic activity (large peaks in $a_p$).

\subsection{Principal component data}
\label{sub:principal_component}

\begin{figure}
    \centering
        \includegraphics{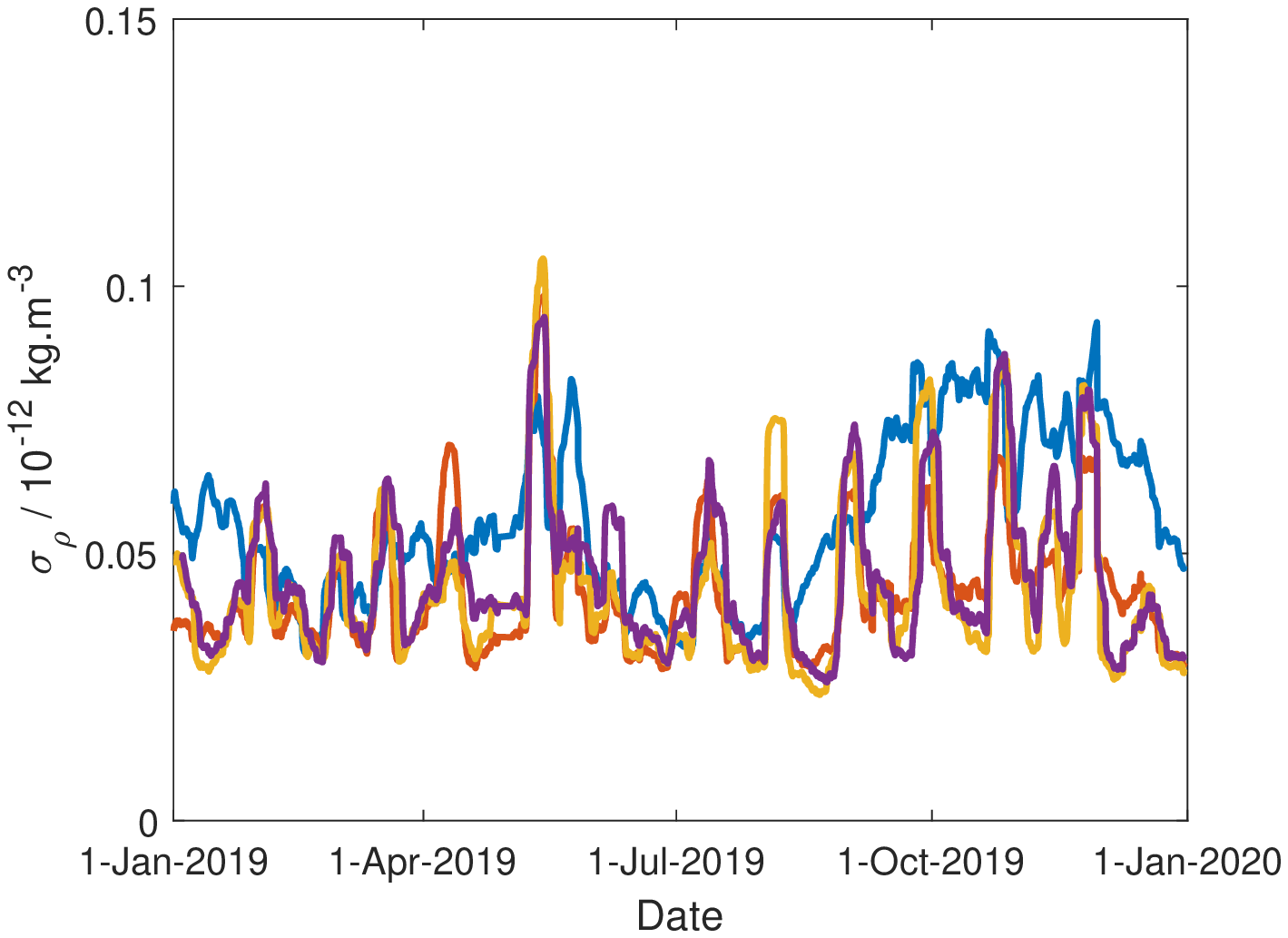}
    \caption{RMS density residual values, taken over a moving window of 100 orbits ($\sim 1$~week). Plotted for NRLMSISE-00 test data linear regression (blue), training data linear regression (orange), 1-day offset Kalman filter calibration (yellow), and 3-day offset Kalman filter calibration (purple) principal component estimates.}\label{fig:movrmsres_pca_msis}
\end{figure}

\begin{figure}
    \centering
        \includegraphics{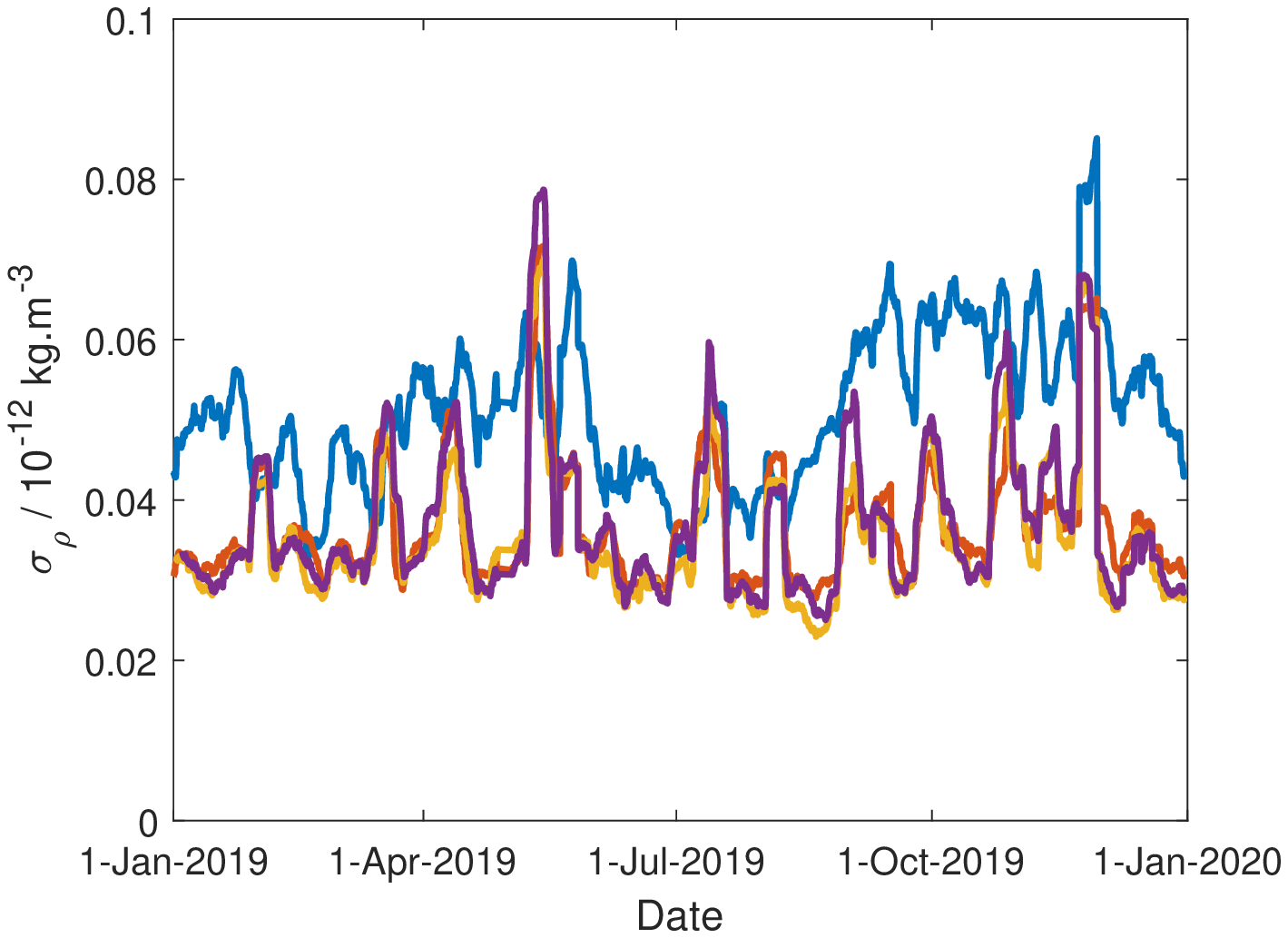}
    \caption{RMS density residual values, taken over a moving window of 100 orbits ($\sim 1$~week). Plotted for combined test data linear regression (blue), training data linear regression (orange), 1-day offset Kalman filter calibration (yellow), and 3-day offset Kalman filter calibration (purple) principal component estimates.}\label{fig:movrmsres_pca_all}
\end{figure}

The calibration technique described in Section~\ref{sec:kalman_filter_approach} was also applied to scores for the first $6$ density principal components. Fixed calibrations of these scores from linear regressions were also computed for comparison. Predicted scores were combined to give density estimates as a function of $u$ over each orbit for each of the empirical models (driven as before). The resulting RMS residuals are tabulated in Tab.~\ref{tab:pca_results}. Temporal variation in density residuals is plotted in Fig.~\ref{fig:movrmsres_pca_msis} for predictions based on NRLMSISE-00 and Fig.~\ref{fig:movrmsres_pca_all} for those based on a combination of model predictions. Similar sharp increases in residuals occurred during the same time periods as those for orbit mean density predictions.

\begin{table}
\centering
\caption{RMS residuals for different density principal component calibration methods. All quantities are shown in $10^{-12}$ kg.m$^{-3}$ (without brackets) and ratio with the mean value (within brackets).}\label{tab:pca_results}
\begin{tabular}{ l c c c c }
\hline
 Calibration method  & NRLMSISE-00 & JB2008 & DTM-2020 & Combined \\
\hline
Linear regression \\ on training data & 0.0581 (0.39) & 0.0508 (0.34) & 0.0582 (0.39) & 0.0521 (0.35) \\
Linear regression \\ on test data & 0.0448 (0.30) & 0.0434 (0.29) & 0.0390 (0.26) & 0.0376 (0.25) \\
$1$-day offset \\ Kalman filter & 0.0457 (0.31) & 0.0431 (0.29) & 0.0382 (0.26) & 0.0364 (0.24) \\
$3$-day offset \\ Kalman filter & 0.0476 (0.32) & 0.0455 (0.30) & 0.0390 (0.26) & 0.0384 (0.26) \\
\hline
\end{tabular}
\end{table}

\section{Discussion}
\label{sec:discussion}

The results presented in Tab.~\ref{tab:oa_results} and Tab.~\ref{tab:pca_results} show that the Kalman filter calibration method provided substantially more accurate predictions than the linear regression on training data. Allowing the relationship between between the thermosphere model output and satellite density measurements to evolve over time helps account for trends not well represented by the climatological and aerodynamic models used. This was true for each of the models examined and their combination estimating the calibration coefficients $1$ and $3$~days ahead.

The proportional improvement using the Kalman filter calibration method was greater for predicting orbit mean densities than densities at specific locations (based on principal component scores). Comparison of Fig.~\ref{fig:movrmsres_oa_msis} and Fig.~\ref{fig:movrmsres_pca_msis} and of Fig.~\ref{fig:movrmsres_oa_all} and Fig.~\ref{fig:movrmsres_pca_all} shows very similar trends in RMS residuals over time. However, there is an approximate floor which is substantially lower for the orbit mean density residuals. The greater improvement in orbit mean density predictions reflects the reduced effect of uncertainty in measurements and effect of unmodelled processes through averaging. Future work will examine the relative importance of these factors.

The Kalman filter calibration method had similar accuracy to applying a constant linear regression across all the test data. Moreover, changes in the accuracy of their predictions over time were highly correlated. Times when residuals were large may indicate when thermosphere variations not well-represented by the empirical models were important.

Changes in the calibration coefficients could have resulted from inaccuracies in thermosphere or satellite aerodynamic modelling. Where the empirical model response to geomagnetic activity or changing solar flux was inaccurate calibration changes would have occurred. Empirical thermosphere models all have greatly reduced performance during geomagnetic storms\cite{Bruinsma2021b}. The combination of any inaccuracy in model local time dependence and nodal precession would have would have produced longer-term changes. Moreover, unmodelled changes in satellite gas-surface interaction properties and thermosphere composition would have caused calibration changes by introducing systematic errors in the POD density measurements. Investigation of the magnitude of such effects would be worthwhile but is beyond the scope of the present study.

The accuracy of this technique may be dependent on where in the solar cycle it is applied. Localised density predictions have previously been reported to be less accurate at solar minimum, suggesting that stochastic effects originating from the lower atmosphere are more prominent at this time \cite{Perez2014}. Moreover, the accuracy of Swarm-C POD thermosphere density measurements is reduced at solar minimum due to relatively low aerodynamic accelerations compared with other non-gravitational forces \cite{VanIJssel2020}. Other satellite-based thermosphere density estimation techniques would have similar issues. Therefore, it is likely that better performance could be obtained applying this technique near solar maximum conditions.

By considering additional data, a better model of the temporal evolution of the calibration coefficients may be obtained. This could involve introducing dependences of $R$ and $\mathbf{M}$ on solar flux and geomagnetic indices. The former would account for change in the magnitude of unmodelled short timescale thermosphere variability and the latter for variation in model calibration. Alternatively, the trivial $\mathbf{F}$ used here could be replaced with some function of these indices.

The methods presented here could form part of a dual Kalman filter approach to orbit determination and orbit localised density estimation. In such an approach one Kalman filter would estimate the orbit state and another would simultaneously estimate the model density calibration. Kalman filtering itself is computationally inexpensive, which would allow such a technique to be implemented aboard a satellite. The approach could be implemented in conjunction with surrogate models representing mean density or principal component scores in terms of geomagnetic and solar flux indices, reducing processing and storage costs associated with density models. The development of operational Kalman filter density prediction tools for satellites would improve the fidelity of orbit determination and prediction. In so doing it would aid accurate and efficient collision avoidance and propulsionless manoeuvring.

\section{Conclusions}
\label{sec:conclusions}
A Kalman filter method has been developed for calibrating thermosphere density predictions from empirical models for a particular satellite. This method substantially improved upon density estimates made using a calibration based on linear regression over previous data. Combination of estimates from different models using a best linear unbiased technique further improved performance. However, the Kalman filter calibration technique suffered from periods of reduced performance, suggestive of short time-scale thermosphere processes which were not well-represented by the climatological empirical models.

\section*{Acknowledgements}
This work was carried out with funding from the Royal Australian Air Force. Resources from the National Computational Infrastructure (NCI), supported by the Australian Government, were used. Thanks go to Dr. Brenton Smith and Dr. Li Qiao for helpful discussions regarding astrodynamics and Kalman filtering. Thanks also to Prof. Russell Boyce and Dr. Melrose Brown for advice and general guidance on preparation of this paper.



 \bibliographystyle{elsarticle-num} 

\begin{thebibliography}{10}
\expandafter\ifx\csname url\endcsname\relax
  \def\url#1{\texttt{#1}}\fi
\expandafter\ifx\csname urlprefix\endcsname\relax\def\urlprefix{URL }\fi
\expandafter\ifx\csname href\endcsname\relax
  \def\href#1#2{#2} \def\path#1{#1}\fi

\bibitem{Vallado2014}
D.~A. Vallado, D.~Finkleman, A critical assessment of satellite drag and
  atmospheric density modeling, Acta Astronautica 95 (2014) 141--165.

\bibitem{Emmert2015}
J.~Emmert, {Thermospheric mass density: A review}, Advances in Space Research
  56~(5) (2015) 773--824.

\bibitem{Stastny2009}
N.~B. Stastny, F.~R. Chavez, C.~Lin, T.~A. Lovell, R.~A. Bettinger, J.~Luck,
  Localized density/drag prediction for improved onboard orbit propagation,
  Tech. rep., AIR FORCE RESEARCH LAB KIRTLAND AFB NM SPACE VEHICLES DIRECTORATE
  (2009).

\bibitem{Perez2014}
D.~P{\'e}rez, B.~Wohlberg, T.~A. Lovell, M.~Shoemaker, R.~Bevilacqua,
  Orbit-centered atmospheric density prediction using artificial neural
  networks, Acta Astronautica 98 (2014) 9--23.

\bibitem{Perez2015}
D.~P{\'e}rez, R.~Bevilacqua, Neural network based calibration of atmospheric
  density models, Acta Astronautica 110 (2015) 58--76.

\bibitem{Friis2008}
E.~Friis-Christensen, H.~L{\"u}hr, D.~Knudsen, R.~Haagmans, Swarm--an earth
  observation mission investigating geospace, Advances in Space Research 41~(1)
  (2008) 210--216.

\bibitem{Mahooti2019}
M.~Mahooti,
  \href{https://www.mathworks.com/matlabcentral/fileexchange/56253-nrlmsise-00-atmosphere-model}{{NRLMSISE-00
  Atmosphere Model}}, [Online; accessed 8-September-2020] (2019).
\newline\urlprefix\url{https://www.mathworks.com/matlabcentral/fileexchange/56253-nrlmsise-00-atmosphere-model}

\bibitem{Mahooti2018}
M.~Mahooti,
  \href{https://www.mathworks.com/matlabcentral/fileexchange/56163-jacchia-bowman-atmospheric-density-model}{{Jacchia-Bowman
  Atmospheric Density Model}}, [Online; accessed 30-April-2021] (2018).
\newline\urlprefix\url{https://www.mathworks.com/matlabcentral/fileexchange/56163-jacchia-bowman-atmospheric-density-model}

\bibitem{SWAMI2021}
D.~Dalaur, S.~Bruinsma, D.~Jackson,
  \href{https://github.com/swami-h2020-eu/mcm}{{SWAMI MCM Model}}, [Online;
  accessed 2-July-2021] (2021).
\newline\urlprefix\url{https://github.com/swami-h2020-eu/mcm}

\bibitem{Picone2002}
J.~Picone, A.~Hedin, D.~P. Drob, A.~Aikin, {NRLMSISE-00 empirical model of the
  atmosphere: Statistical comparisons and scientific issues}, Journal of
  Geophysical Research: Space Physics 107~(A12) (2002) SIA--15.

\bibitem{Hedin1977a}
A.~Hedin, J.~Salah, J.~Evans, C.~Reber, G.~Newton, N.~Spencer, D.~C. Kayser,
  D.~Alcayde, P.~Bauer, L.~Cogger, et~al., {A global thermospheric model based
  on mass spectrometer and incoherent scatter data MSIS, 1. N2 density and
  temperature}, Journal of Geophysical Research 82~(16) (1977) 2139--2147.

\bibitem{Hedin1977b}
A.~Hedin, C.~Reber, G.~Newton, N.~Spencer, H.~Brinton, H.~Mayr, W.~Potter, {A
  global thermospheric model based on mass spectrometer and incoherent scatter
  data MSIS, 2. Composition}, Journal of Geophysical Research 82~(16) (1977)
  2148--2156.

\bibitem{Bowman2008}
B.~Bowman, W.~K. Tobiska, F.~Marcos, C.~Huang, C.~Lin, W.~Burke, {A new
  empirical thermospheric density model JB2008 using new solar and geomagnetic
  indices}, in: AIAA/AAS astrodynamics specialist conference and exhibit, 2008,
  p. 6438.

\bibitem{Bruinsma2021a}
S.~Bruinsma, {The DTM2020 models}, in: 43rd COSPAR Scientific Assembly. Held 28
  January-4 February, Vol.~43, 2021, p. 813.

\bibitem{Barlier1978}
F.~Barlier, C.~Berger, J.~Falin, G.~Kockarts, G.~Thuillier, A thermospheric
  model based on satellite drag data., in: Annales de Geophysique, Vol.~34,
  1978, pp. 9--24.

\bibitem{Bruinsma2015}
S.~Bruinsma, {The DTM-2013 thermosphere model}, Journal of Space Weather and
  Space Climate 5 (2015) A1.

\bibitem{Kalman1960}
R.~E. Kalman, A new approach to linear filtering and prediction problems,
  Transaction of the ASME-Journal of Basic Engineering (1960) 35--45.

\bibitem{Mehra1972}
R.~Mehra, Approaches to adaptive filtering, IEEE Transactions on automatic
  control 17~(5) (1972) 693--698.

\bibitem{Pinheiro1996}
J.~C. Pinheiro, D.~M. Bates, Unconstrained parametrizations for
  variance-covariance matrices, Statistics and computing 6~(3) (1996) 289--296.

\bibitem{lyons1988}
L.~Lyons, D.~Gibaut, P.~Clifford, How to combine correlated estimates of a
  single physical quantity, Nuclear Instruments and Methods in Physics Research
  Section A: Accelerators, Spectrometers, Detectors and Associated Equipment
  270~(1) (1988) 110--117.

\bibitem{VanIJssel2020}
J.~van~den IJssel, E.~Doornbos, E.~Iorfida, G.~March, C.~Siemes,
  O.~Montenbruck, Thermosphere densities derived from swarm gps observations,
  Advances in Space Research 65~(7) (2020) 1758--1771.

\bibitem{CLS}
{Collecte Localisation Satellites}, \href{https://spaceweather.cls.fr/}{Space
  weather services at {CLS}} (2015).
\newline\urlprefix\url{https://spaceweather.cls.fr/}

\bibitem{GFZ}
{GFZ German Research Centre for Geosciences},
  \href{https://www.gfz-potsdam.de/en/kp-index/}{{Geomagnetic Kp Index}}
  (2020).
\newline\urlprefix\url{https://www.gfz-potsdam.de/en/kp-index/}

\bibitem{SET}
{Space Environment Technologies},
  \href{https://sol.spacenvironment.net/JB2008/index.php}{{JB2008}} (2021).
\newline\urlprefix\url{https://sol.spacenvironment.net/JB2008/index.php}

\bibitem{Bruinsma2021b}
S.~Bruinsma, C.~Boniface, E.~K. Sutton, M.~Fedrizzi, Thermosphere modeling
  capabilities assessment: geomagnetic storms, Journal of Space Weather and
  Space Climate 11 (2021) 12.

\end{thebibliography}


\end{document}